\documentclass[letterpaper, 10 pt, conference]{ieeeconf}  

\IEEEoverridecommandlockouts                              

\usepackage{graphicx}
\usepackage{balance}
\usepackage{comment}
\usepackage{cite}
\usepackage{amssymb}
\usepackage[tight,footnotesize]{subfigure}
\usepackage[active]{srcltx}
\usepackage{amsmath}
\usepackage{amsfonts}

\newcommand{\norm}[1]{\left\lVert#1\right\rVert}
\graphicspath{{./figs/}}
\usepackage{color,soul}
\usepackage{eurosym}
\usepackage{algorithm}
\usepackage{algorithmic}
\usepackage{multicol}
\usepackage{xcolor}
\UseRawInputEncoding
\usepackage{esvect}
\usepackage{bm}
\usepackage{dsfont}

\title{\LARGE \bf
Downing a Rogue Drone with a Team of Aerial Radio Signal Jammers
}

\author{Savvas~Papaioannou,~Panayiotis~Kolios~and Georgios Ellinas 
\thanks{The authors are with the KIOS Research and Innovation Centre of Excellence (KIOS CoE) and the Department of Electrical and Computer Engineering, University of Cyprus, Nicosia, 1678, Cyprus. E-mail:{\tt\small \{papaioannou.savvas, pkolios, gellinas\}@ucy.ac.cy}%
\newline
This work is supported by the European Union's Horizon 2020 research and innovation programme under grant agreement No 739551 (KIOS CoE) and from the Republic of Cyprus through the Directorate General for European Programmes, Coordination and Development.}}

\begin{document}

\maketitle

\begin{abstract}
This work proposes a novel distributed control framework in which a team of pursuer agents equipped with a radio jamming device cooperate in order to track and radio-jam a rogue target in 3D space, with the ultimate purpose of disrupting its communication and navigation circuitry. The target evolves in 3D space according to a stochastic dynamical model and it can appear and disappear from the surveillance area at random times. The pursuer agents cooperate in order to estimate the probability of target existence and its spatial density from a set of noisy measurements in the presence of clutter. Additionally, the proposed control framework allows a team of pursuer agents to optimally choose their radio transmission levels and their mobility control actions in order to ensure uninterrupted radio jamming to the target, as well as to avoid the jamming interference among the team of pursuer agents. Extensive simulation analysis of the system's performance validates the applicability of the proposed approach. 
\end{abstract}

\section{Introduction} \label{sec:intro}

Drones which have nowadays become the new emerging trend, are currently being utilized in many applications ranging from aerial photography and critical infrastructure inspection to emergency-response missions and aerial monitoring tasks. Unfortunately, drones have also become a threat and a risk to public safety. In particular, numerous times drones have threatened public safety by targeting airports and restricted airspaces \cite{Schneider2019}, attacking critical infrastructures \cite{AC1} and endangering people's lives \cite{Humphreys2015}.
For this reason there is a necessity for counter-drone systems that can detect, track, and interdict rogue or malicious drones \cite{Wesson2013}. Although, counter-drone approaches and systems have already been proposed in the literature \cite{Guvenc2018,Loeb2017,Shi2018}, these are still in their infancy and considerably more work is needed in order for this technology to reach the required level of maturity. 

In this work, a distributed multi-agent control framework is proposed, in which a team of pursuer agents (i.e., pursuer drones) cooperate in order to continuously track and radio-jam a target (i.e., rogue drone) in 3D space (as depicted in Fig. \ref{fig:problem}), disrupting its communication and sensing circuitry and thus ultimately forcing it to execute its fail-safe protocols \cite{Revill2016} i.e., auto-landing. The assumption is that the pursuer agents are equipped with a 3D range-finding active sensor \cite{Park2001} with limited sensing range, which they use to obtain noisy target measurements (i.e., radial distance, azimuth angle, and inclination angle) in the presence of false-alarm measurements (or clutter). In this work it is assumed that the target detection process is uncertain and that the target can appear and disappear within the surveillance area at random times (i.e., the target can spawn from anywhere inside the surveillance area and additionally during tracking it can move behind obstacles or other occlusions which results in its disappearance from the field of view of the pursuer agents). Finally, it is assumed that the pursuer agents have the ability to transmit power to the target via their on-board active sensor, at discrete power-levels, with the ultimate purpose of radio-jamming its circuitry. 
\noindent The main contributions of this work are two-fold:

\begin{figure}
	\centering
	\includegraphics[width=\columnwidth]{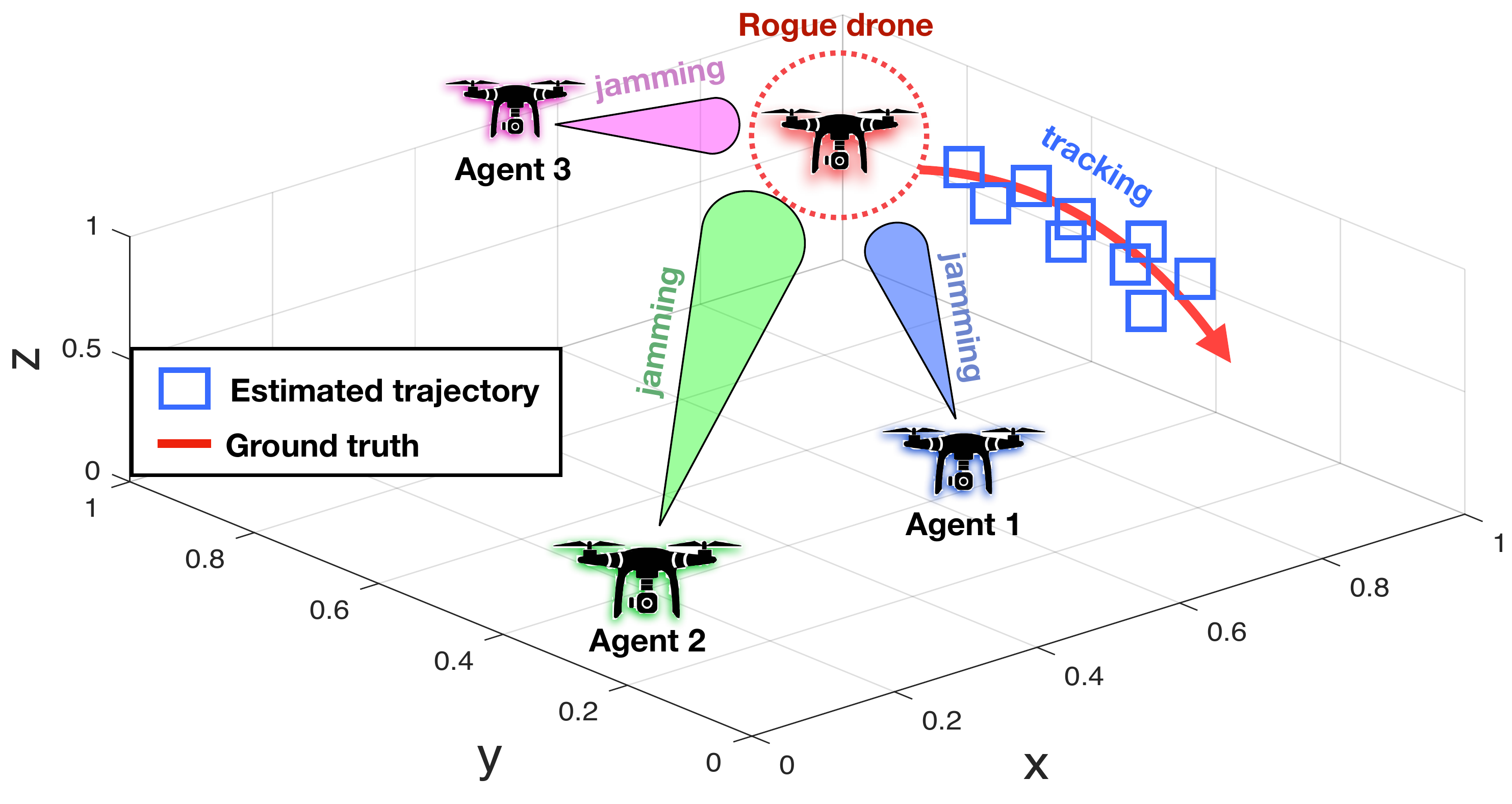}
	\caption{A team of pursuer agents choose their mobility controls and radio transmission levels which result in accurate tracking and uninterrupted radio jamming of the roque target.}	
	\label{fig:problem}
	\vspace{-4 mm}
\end{figure}

\begin{itemize}
    \item A novel distributed control framework is proposed for the problem of target tracking and target radio-jamming in 3D by a team of cooperative mobile agents. The proposed approach allows the team of pursuer agents to optimally (a) choose their mobility control actions that result in accurate target tracking and (b) choose their transmit power levels to cause uninterrupted target radio jamming. The agents cooperate for improving the target tracking-and-jamming performance of the team while minimizing the jamming interference amongst them.
    \item In the scenario considered, the target can exist in one of two states i.e., present or absent. Thus in order to be jammed, its existence probability along with its spatial density must be estimated from a sequence of noisy measurements in the presence of clutter, by a team of mobile agents equipped with conic directional antennas with limited sensing range.
    \end{itemize}

\noindent The rest of the paper is organized as follows. Section \ref{sec:related_work} discusses the related work. Section \ref{sec:system_overview} formulates the problem and illustrates the proposed system architecture. Section \ref{sec:system_model} develops the system model and Section \ref{sec:proposed_approach} discuses the details of the proposed approach. Finally, Section \ref{sec:Evaluation} conducts an extensive performance evaluation and Section \ref{sec:Conclusion} concludes the paper and discusses future directions.

\section{Related Work} \label{sec:related_work}

A recent survey on detection, tracking, and interdiction techniques for malicious drones or unmanned aerial vehicles (UAVs) can be found in \cite{Guvenc2018}, where the authors discuss various drone threats and review several counter-measures that have been investigated in the literature. In \cite{Shi2018} the authors present a detailed survey on the state-of-the-art anti-drone systems and discuss techniques and technologies used for drone surveillance. The work in \cite{Ganti2016} discusses the advantages and disadvantages of various drone detection methods including audio-visual, thermal, and RF and proposes a low-cost stationary system for drone detection and tracking which can be easily incorporated into third-party anti-drone platforms. The authors in \cite{Srigrarom2020} focus on the problem of detection and tracking of small and fast moving drones with the ultimate goal of developing a system that can be used to prevent such small drones from accessing restricted areas and facilities. 

Moreover, in \cite{Karas2020,Papaioannou2019_3,Papaioannou2019_2,Papaioannou2019_1,Papaioannou4}, the problems of formation control, target tracking, and target interception with multiple agents are investigated but without considering jamming capabilities for the pursuers. The work in \cite{Perkins2015} develops a UAV based solution for localizing a GPS jammer and the  work in \cite{Multerer2017} proposes a low-cost ground jamming system to counteract the operation of small drones. In \cite{Brust2017} a team of defense agents forms a cluster around a malicious target in order to prevent it from entering restricted airspaces. The authors assume however the availability of a tracking system for detecting and tracking the target. Finally, in \cite{Papaioannou2019,Valianti} the problem of jamming a rogue drone with a team of mobile agents is investigated, however without  considering target appearance/disappearance events. 

Compared to the existing techniques, the proposed framework develops a novel cooperative technique for the problem of target tracking and target radio-jamming in 3D by a team of mobile agents, while considering the induced jamming interference amongst the team. The problem is investigated in the presence of target detection uncertainty, clutter, and target appearance/disappearance events which to the best of our knowledge has not been investigated before.

\begin{figure}
	\centering
	\includegraphics[width=\columnwidth]{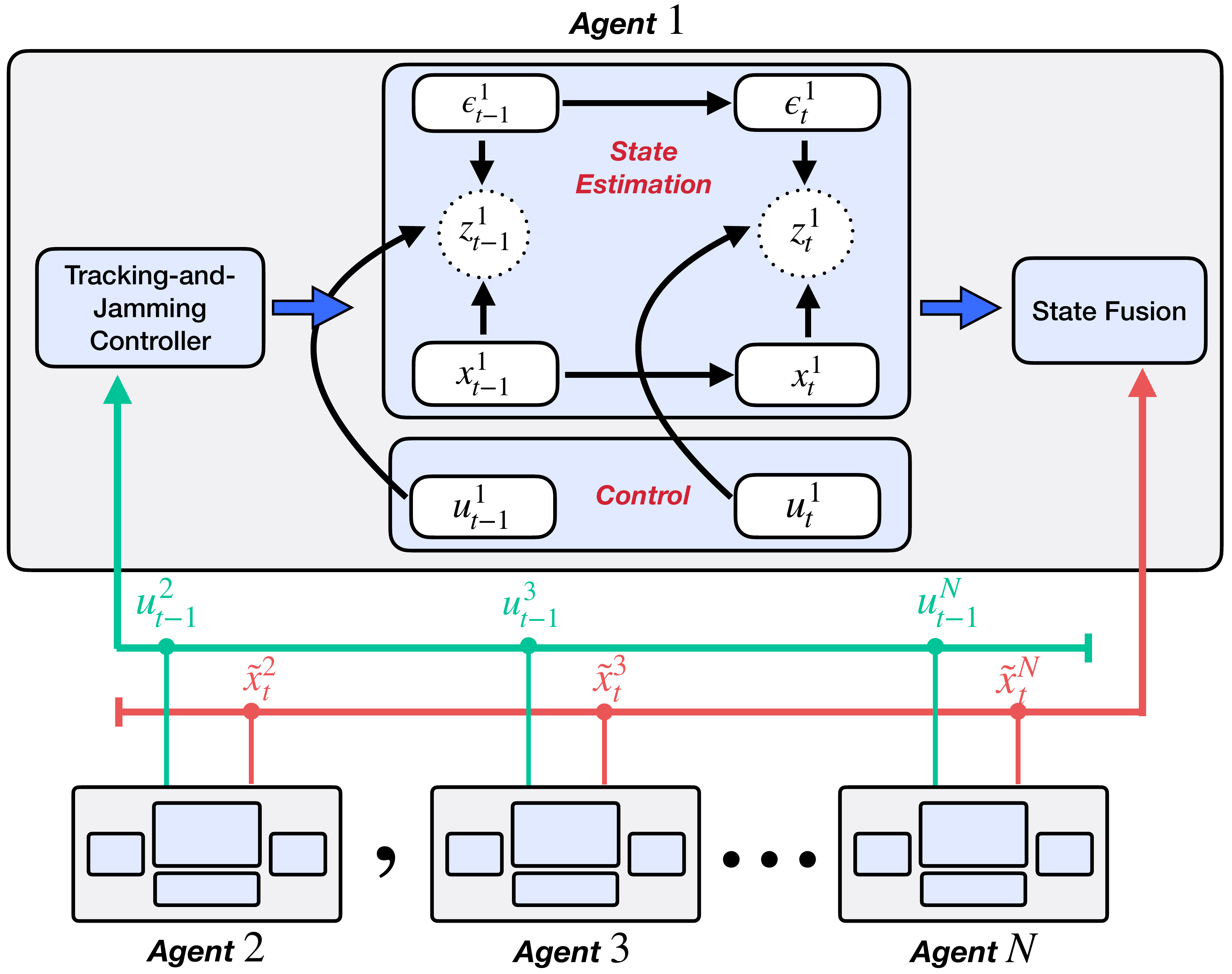}
	\caption{Proposed system architecture. The symbols denote $\epsilon_t:$ target existence, $x_t:$ target state, $z_t:$ target measurement and $u_t:$ agent controls.}	
	\label{fig:sys_arch}
	\vspace{-5mm}
\end{figure}

\section{System Overview}\label{sec:system_overview}
The problem tackled in this work can be stated as follows: \textit {At each time-step $t$ each agent $j \in [1,..,N]$ must decide its mobility control action $u_t^{j} \in \mathbb{U}_t^j$ and transmit power-level $\ell^j_t \in \mathbb{L}_t^j$ that results in accurate target tracking and uninterrupted target radio-jamming. At each time-step the agents cooperate to maximize the joint tracking and radio-jamming performance of the system, while avoiding the induced radio-jamming interference amongst them}.

Figure \ref{fig:sys_arch} illustrates the proposed system architecture. 
Each agent $j \in [1,..,N]$ uses stochastic filtering in order to estimate at each time-step $t$: (a) the probability mass function of the target existence $p^j_{t}(\epsilon^j_t|Z_{1:t})$ given target measurements up to time $t$, $Z^j_{1:t}$ and (b) the filtering distribution $f_{t}^j(x_t|Z_{1:t})$ of the target state $x^j_t$. This is illustrated by the probabilistic graphical model in Fig. \ref{fig:sys_arch}, where the target state $x^j_t$ and the target existence $\epsilon^j_t$ are the random variables of an unobserved Markov process with observations $z^j_t$ (i.e., target measurement). In the problem considered, the target measurement $z^j_t$ received by agent $j$ depends on the agent's applied mobility control action $u^j_t$ as shown in the figure. As a result, to optimally estimate the state of the target, the agent must make optimized mobility control decisions which is achieved by the proposed controller. The objective of the \textit{tracking-and-jamming} controller is not only to find the optimal mobility control actions $u^j_t$  but also the optimal transmit power-level $\ell^j_t$ for agent $j$ such that the  received power at the rogue drone is maximized in order to achieve uninterrupted radio jamming. The agents cooperate by exchanging information (i.e., their current state and target estimated state) in order to optimize the target radio-jamming and avoid the jamming interference amongst them. This work does not consider a target that has the potential to jam the pursuer agents. Moreover, it is assumed that each agent is able to identify with absolute certainty if another entity is a pursuer agent, i.e., an agent cannot be mistaken for a target by another pursuer agent.

\section{System Model} \label{sec:system_model}

\subsection{Target Dynamics}\label{ssec:target_dynamics}
This work considers one target (e.g., a rogue drone) which can appear/disappear at random times anywhere inside the surveillance area and thus at any time-step the target can exist in one of two states i.e., present or absent. When the target is present, it moves in 3D space according to a stochastic dynamical model. More specifically, the state of the target at time $t$ is modeled as a Bernoulli random finite set (RFS) \cite{VoBook2015,Mahler2014book} $X_t \in \{\emptyset,\{x_t\}\}$, with Bernoulli RFS \cite{Ristic2013} density $F_{t}(X_t)$ with parameters $(p_{t}(\epsilon_t|Z_{1:t}), f_{t}(x_t|Z_{1:t}))$, where $e_{t}=p_{t}(\epsilon_t=1|Z_{1:t})$ is the probability of target existence given measurements $Z_{1:t}$ up to time $t$ and $s_{t} = f_{t}(x_t|Z_{1:t})$ is the spatial density of the target with state $x_t$. The dynamics of the target are modeled as a Bernoulli Markov Process \cite{Mahler2007book} with transitional RFS density $\psi_{t|t-1}(X_t|X_{t-1})$ given by:

\vspace{+3mm}
\begin{tabular}{ l|c|c| }
\multicolumn{1}{r}{}
 &  \multicolumn{1}{c}{$X_t =\emptyset$}
 & \multicolumn{1}{c}{$X_t =\{x_t\}$} \\
\cline{2-3}
$X_{t-1} =\emptyset$ & $1-p_b$ & $p_b b_t(x_t)$ \\
\cline{2-3}
$X_{t-1} =\{x_{t-1}\}$& $1-p_s$ & $p_s \pi_{t|t-1}(x_t|x_{t-1})$  \\
\cline{2-3}
\end{tabular}
\vspace{+3mm}   

\noindent where $p_b$ denotes the probability of target birth, $b_t(x)$ is the birth density (uniform inside the surveillance area), $p_s$ is the probability that the target survives to the next time-step, and $\pi_{t|t-1}(x_t|x_{t-1})$ is the target transitional density which in this work it is assumed to be governed by the following discrete-time dynamical model:
\begin{equation} \label{eq:target_dynamics}
    x_t = \Phi x_{t-1} + \Gamma \nu_{t}
\end{equation}
where $x_t = [\text{x},\text{y},\text{z},\dot{\text{x}},\dot{\text{y}},\dot{\text{z}}]_t^\top \in \mathcal{X}$ denotes the target state at time $t$ which consists of the position and velocity components in 3D Cartesian coordinates and $\nu_{t} \sim \mathcal{N}(0,\Sigma_\text{v})$ denotes the perturbing acceleration noise which is drawn from a zero mean multivariate normal distribution with covariance matrix $\Sigma_\text{v}$. The matrices $\Phi$ and $\Gamma$ are defined as:
\begin{equation}
\Phi = 
\begin{bmatrix}
    \text{I}_3 & \Delta T \cdot \text{I}_3\\
    \text{0}_3 & \text{I}_3
   \end{bmatrix},
\Gamma = 
\begin{bmatrix}
    0.5\Delta T^2 \cdot \text{I}_3\\
     \Delta T \cdot \text{I}_3
   \end{bmatrix}
\end{equation}

\noindent where $\Delta T$ is the sampling period and $\text{I}_3$, $\text{0}_3$ are the $3 \times 3$ identity and zero matrix, respectively.

\subsection{Pursuer Agent Kinematics} \label{ssec:AgentDynamics}
A team of $N$ autonomous pursuer agents (e.g., UAV agents) operate inside the surveillance area,  each of which is subject to the following kinematic model:

\begin{equation} \label{eq:controlVectors}
u^j_{t}\!\! = \! u^j_{t-1} + \begin{bmatrix}
						\Delta_R(l_1) \sin(l_2 \Delta_\phi) \cos(l_3 \Delta_\theta)\\
						\Delta_R(l_1) \sin(l_2 \Delta_\phi) \sin(l_3 \Delta_\theta)\\
						\Delta_R(l_1) \cos(l_2 \Delta_\phi)
					\end{bmatrix}\!\!\!,\!\!  
					\begin{array}{l} 
					    l_1 = 1,...,|\Delta_R|\\
					    l_2 = 0,...,N_\phi\\ 
						l_3 = 1,...,N_\theta
				    \end{array} 
\end{equation}
where $j \in [1,..,N]$, $u^j_{t} = [u^j_x,u^j_y,u^j_z]^\top_{t} \in \mathbb{R}^3$ denotes the state (i.e., position) of pursuer agent $j$ at time $t$, $\Delta_R$ is a vector of possible radial step sizes, $\Delta_\phi=\pi/N_\phi$, $\Delta_\theta=2\pi/N_\theta$, and the parameters $(|\Delta_R|,N_\phi,N_\theta)$ determine the number of possible mobility control actions. The set of all admissible control actions of agent $j$ at time $t$ is denoted as $\mathbb{U}^j_{t}=\{u^{j,1}_{t},u^{j,2}_{t},...,u^{j,|\mathbb{U}^j_{t}|}_{t}  \}$ according to Eq. (\ref{eq:controlVectors}). Although in this work a simplified kinematic model for the agents is utilized in order to demonstrate the proposed approach, depending on the application scenario more realistic kinematic/dynamic models can also be incorporated.

\subsection{Agent Sensing and Jamming Model} \label{ssec:agent_sensing}
The pursuer agents are equipped with an onboard directional active 3D range-finding radio which they use in order to detect the target, acquire target measurements, and transmit power to the target in order to radio-jam its communications circuitry. The range-finding characteristics are as follows:

\subsubsection{Sensing Profile} The radio's sensing profile $S$ in 3D is modeled as a circular right angle cone with Cartesian coordinates $(x,y,z)$ given by: $[ x=u\cos(v), y=u\sin(v), z=\frac{h_a}{r} u ]$ where $ u \in [0, r], ~ v \in [0, 2\pi]$, $h_a$ characterizes the effective sensing range, $r=\tan(\frac{\theta_a}{2})h_a$ is the base radius of the cone, and $\theta_a$ is the opening angle of the cone. Thus, a target with position coordinates $Hx_t = [\text{x},\text{y},\text{z}]_t^\top$ (where $H$ is a matrix which extracts the position coordinates from state $x_t$) resides inside agent's $j$ sensing range when $Hx_t \in S^j$. It should be mentioned that the direction of the agent's sensor is given at each time-step by the direction of the vector $\vec{d}_t = x_t - u^j_t$, where $x_t$ and $u^j_t$ are the positions of the target and  agent $j$, respectively.

\subsubsection{Measurement Model} Each agent $j$ uses its radio received signal to acquire 3D target measurements $z^j_t=[\rho,\theta,\phi]^\top_t \in \mathcal{Z}$ (i.e., radial distance $\rho$, azimuth angle $\theta$, and inclination angle $\phi$) according to the following measurement model: 

\begin{equation}
    z^j_t = h(x_t,u^j_t) + w^j_t
\end{equation}
The function $h(x_t,u^j_t)$ is given by:
\begin{equation}
    \left[ \eta_t, ~\tan^{-1}\left(\frac{\Delta_{t,y}}{\Delta_{t,x}}\right),~ \tan^{-1}\left(\frac{\sqrt{\Delta_{t,x}^2+\Delta_{t,y}^2}}{\Delta_{t,z}}\right) \right]^\top
\end{equation} 
where $\eta^j_t=\norm{H x_t-u^j_t}_2$, $\Delta_{t,y} = [\text{y}-u^j_y]_t$, $\Delta_{t,x} = [\text{x}-u^j_x]_t$, $\Delta_{t,z} = [\text{z}-u^j_z]_t$, and $w^j_t \sim \mathcal{N}(0,\Sigma_w)$ is zero mean Gaussian measurement noise with covariance matrix $\Sigma_w$. 
Due to sensing imperfections, at time $t$, agent $j$ also receives false-alarm measurements or clutter $\{c^{j1}_t,\ldots,c^{jn}_t\}$ (in addition to the target measurement) with a Poisson rate of $\lambda_c$ (i.e., $\mathbb{E}(n) = \lambda_c$), which are uniformly distributed (i.e., with clutter density denoted as $p_c(c_t)$) inside the measurement space. To summarize, at each time-step, agent $j$ receives a set $Y^j_t$ of measurements which is given by: 
\begin{equation}\label{eq:Upsilon}
    Y^j_t=\bigcup \Big\{ y_1\subset\{ \emptyset,z^j_t\}, y_2 \subseteq \{c^{j1}_t,\ldots,c^{jn}_t\} \Big\}
\end{equation}
\noindent Thus, agent $j$ can receive at time-step $t$, zero or one target measurements $z^j_t$ and a set of false-alarms measurements.

\subsubsection{Target Detection and Jamming}
The agents use their onboard radio to detect and jam the target. A target detection occurs, with certain probability, when the target is illuminated (i.e., when the agent transmits power). Hence, it is important to note here that the same transmit signals are used to detect the target and at the same time jam it. The detection probability i.e., $p^j_D = p^j_D(x_t,u^j_t,l^j_t)$, that agent $j$ with state $u^j_t$ and transmit power-level $l^j_t$ detects a target with state $x_t$ is given by: 
\begin{equation}\label{eq:sensing_model}
 p^j_D(x_t,u^j_t,l^j_t) = 
  \begin{cases} 
   \frac{l^j_t}{l^j_{\max}} p^{\max}_D & \!\!\!\!\!\!\text{if } \eta^j_t < R_0 \wedge x_t \in S^j \\
   \frac{l^j_t}{l^j_{\max}} p^{\max}_D  \left(\frac{R_0}{\eta_t}\right)^{n_e} &\!\!\!\!\!\! \text{if } \eta_t > R_0 \wedge x_t \in S^j \notag
  \end{cases}
\end{equation}

\noindent where $l^j_{\max}$ is the maximum transmit power-level, $\eta^j_t=\norm{H x_t-u^j_t}_2$ denotes the Euclidean distance between the agent and the target in 3D space, $p^{\max}_D$ denotes the maximum attainable detection probability of the sensor which can be obtained when a target resides within $R_0$ distance from the agent's position and the target also resides inside the agent's sensing profile $S^j$, and $n_e$ is the path-loss exponent. Note that, $p^j_D = 0$, when $x_t \notin S^j$.
Finally, the transmit power-level $l^j_t$ takes its values from the discrete set of admissible transmit power-levels $\mathbb{L}^j$, i.e., $l^j_t \in \mathbb{L}^j = \{l^{j1}_t,...,l^{jn}_t \}$ with $l^j_{\max} = \max ~\mathbb{L}^j$. Additionally, the received power at the target with location $Hx_t$ from an agent with state $u^j_t$ which transmits at power-level $l^j_t$ is given by the following path-loss model:
\begin{equation}\label{eq:pathLoss}
R^j_t(Hx_t,u^j_t,l^j_t) = 
\begin{cases} 
    l^j_t & \text{if } \eta^j_t < R_0 \wedge x_t \in S^j \\
    l^j_t  \left(\frac{R_0}{\eta_t}\right)^{n_e} & \text{if } \eta_t > R_0 \wedge x_t \in S^j \notag\\
\end{cases} 
\end{equation}
Note that, $R^j_t(Hx_t,u^j_t,l^j_t) = 0$, when $x_t \notin S^j$.

\section{Cooperative Tracking and Jamming} \label{sec:proposed_approach}

The \textit{tracking-and-jamming} control module depicted in Fig. \ref{fig:sys_arch} seeks to find the joint mobility and power-level control actions that result in uninterrupted target tracking and target radio-jamming. This section discusses the details of the proposed control approach.

\subsection{Target State Estimation} \label{ssec:state_estimation}
Each agent $j$ estimates the probability of target existence $e^j_{t}$ and the target spatial density $s^j_{t}$ by maintaining and propagating in time the multi-object probability distribution $F^j_{t}(X^j_t|Y^j_{1:t})$ of the RFS target state $X^j_t$ given measurements $Y^j_{1:t} = \{Y^j_1,..,Y^j_t\}$, using multi-object stochastic filtering \cite{Mahler2014book} as shown below (as a note, the index on the agent is dropped for notational clarity):
\begin{align} 
& \!\!\!\!\!\!\!\!\!\!\!\!\!\!\!\!\!\!\!\!\!\!\!\!\!\!\!\!\!\! F_{t|t-1}(X_t|Y_{1:t-1}) =  \label{eq:predict}\\
& \!\!\!\!\!\!\!\!\!\! \int  \psi_{t|t-1}(X_t|X_{t-1})  F_{t-1}(X_{t-1}|Y_{1:t-1}) d X_{t-1} \notag \\
\!\!\!\!\!\!\!\!\!\! F_{t}(X_t|Y_{1:t}) &= \frac{\phi_t(Y_t|X_t)  F_{t|t-1}(X_t|Y_{1:t-1})}{\int \phi_t(Y_t|X_t) F_{t|t-1}(X_t|Y_{1:t-1}) dX_t} \label{eq:update}
\end{align}

\noindent where $\psi_{t|t-1}(X_t|X_{t-1})$ is the target RFS transitional density as discussed in Section \ref{ssec:target_dynamics}, $\phi_t(Y_t|X_t)$ is the RFS measurement likelihood function according to the measurement model discussed in Section \ref{ssec:agent_sensing}, $F_{t|t-1}(X_t|Y_{1:t-1})$ is the multi-object predictive density, and finally $F_{t}(X_t|Y_{1:t})$ is the multi-object posterior filtering density. The recursion shown here is a generalization of the Bayes filter \cite{Sarkka2013} on random finite sets \cite{Mahler2007book}. The solution of the above recursion  (i.e., see \cite{Ristic2013}) for the modeling assumptions discussed in Section \ref{sec:system_model} allows each agent to compute recursively (over time) $e_{t}$ and $s_{t}$ via the Bernoulli filter recursion \cite{Ristic2013} as:
\begin{align}
    e_{t|t-1} &= p_b(1-e_{t-1}) + p_s e_{t-1} \notag  \\
    s_{t|t-1}\!(x_t)\! &= \!q_{t|t-1}\!\! +\! \frac{p_s e_{t-1} \!\int\!\! \pi_{t|t-1}\!(x_t|x_{t-1})s_{t-1}\!(x_{t-1}) dx_{t-1} }{e_{t|t-1}} \notag
\end{align}
for the prediction step and
\begin{align}
    e_{t} &= \frac{1 - q_t}{1 - e_{t|t-1} q_t} e_{t|t-1} \notag \\
    s_{t}(x_t) &= \frac{(1-p_D) + p_D \sum_{y\in Y_t} \frac{g_t(y|x_t,u_t)}{\lambda_c p_c(y)}}{1-q_t} s_{t|t-1}(x_t) \notag
\end{align}
for the update step, where $p_D(x_t,u_t,l_t)$ is abbreviated as $p_D$, $q_{t|t-1}$ and $q_t$ are defined as $q_{t|t-1} = p_b (1-e_{t-1})b_t(x_t) e_{t|t-1}^{-1}$, $q_t = \int p_D s_{t|t-1}(x) dx - \sum_{y \in Y_t} \int p_D g_t(y|x,u_t) s_{t|t-1}(x) dx \left(\lambda_c p_c(y) \right)^{-1}$, and finally $g_t(y|x_t,u_t) = \mathcal{N}(y; h(x_t,u_t),\Sigma_w)$ is the measurement likelihood function for the target generated measurement.

That said, each agent $j$ computes at each time-step $e^j_{t}$ and $s^j_{t}$ using the recursion shown above. The final target state is then fused as follows: The target existence probability is first exchanged among all agents and the fused estimate is computed as  $\hat{e}_{t} = \text{mean} ~\{ e^j_{t}~ |~ j=1,..,N\}$. Then, the set of agents that sense the presence of the target i.e., $\tilde{N}=\{j ~|~ e^j_{t} > 0.5\}$ first extract the estimated target state $\hat{x}^j_t$ from $s^j_{t}$ and its covariance matrix $C^j$. Then $[ \hat{x}^j_t, C^j],~\forall j \in \tilde{N} $ is exchanged among all $\tilde{N}$ agents and combined using covariance intersection \cite{Julier2017,Deng2012}. Finally, the fused results are communicated to all $N$ agents which they use to sample from and update their filtering densities.

\subsection{Single Agent Tracking-and-Jamming Control} \label{ssec:single_control}
To achieve tracking-and-jamming control first observe that the posterior target existence probability $e_{t}$, the posterior filtering density $s_{t}$, and the target received power $R_t$ are all influenced by the target detection events. Thus, optimizing $u_t$ and $l_t$ for maximum target detection performance, results in optimized tracking-and-jamming performance. Optimizing the target detection, results in more accurate estimation of $e_{t}$ and $s_{t}$, from which the target state can be extracted and thus target radio jamming can be maximized. Hence, the optimal control actions $(\hat{u}^j_t,\hat{l}^j_t)$ for agent $j$ can be computed as: 
\begin{equation}
    \hat{u}_{t}^{j},\hat{l}^j_t = \underset{u^j_{t} \in \mathbb{U}^j_{t},l^j_{t} \in \mathbb{L}^j_{t}}{\arg\max}~ p^j_D(\hat{x}_t,u^j_t,l^j_t)
\end{equation}
where $\hat{x}_t$ is the estimated state of the target. It should be noted, however, that $\hat{x}_t$ is not available until the control actions $\hat{u}^j_t$ and $\hat{l}^j_t$  have already been chosen and applied. In order to bypass this problem, each agent $j$ approximates $\hat{x}^j_t \sim \tilde{x}^j_t$ as:
\begin{equation}
    \tilde{x}^j_{t} = \int x_t s^j_{t|t-1}(x_t) d x_t, ~\text{iff}~ e^j_{t|t-1} > 0.5
\end{equation}
where $s^j_{t|t-1}(x_t)$ is the predictive spatial density of the target and $e^j_{t|t-1}$ is the predicted probability of target existence.


\subsection{Centralized Tracking-and-Jamming Control}
To tackle the cooperative tracking-and-jamming control problem, the induced jamming interference among the agents must be kept below the critical value so that the agents can remain operational at all times. The joint target detection probability for which exactly $m$ out of $N$ agents jointly track-and-jam the target is given by:
\begin{equation}
\xi^N_m = \underset{1 \le i_1 < i_2 < ,..., < i_m \le N }{\sum}~ \prod_{j=1}^{N}~ \frac{(1-p^j_D) ~p^{i_j}_D}{1-p^{i_j}_D}
\end{equation}

\noindent where $p^j_D(\bar{x}_t,u^j_t,l^j_t)$ is abbreviated as $p^j_D$ for notational clarity, and $\bar{x}_t = \text{mean}\{\tilde{x}^j_{t}: \forall j\}$ denotes the mean predicted target state from all agents $j$ with $e^j_{t|t-1} > 0.5$. The operator $\sum_{1 \le i_1 < i_2 < ,..., < i_m \le N }(.)$ computes the combinations ${N \choose m}$ among the agents. To ensure that at least $n$ out of $N$ agents effectively track-and-jam the target, while at the same time respect the interference constraints amongst them, the cooperative tracking-and-jamming control objective is defined as $\Xi^N_n = \sum_{m=n}^N \xi^N_m$ and the problem becomes:

\begin{subequations}
\begin{align} 
     (\hat{u}^j_t,\hat{l}^j_t) &= \underset{u^j_{t} \in \mathbb{U}^j_{t},l^j_{t} \in \mathbb{L}^j_{t}}{\arg\max} ~\Xi^N_n, ~ \forall j \in [1,..,N] \label{eq:joint_control1}\\
    \text{s.t.} &~ \sum_{j=1}^N R^j_t(u^{i \ne j}_t,u^j_t,l^j_t) < \delta^i,~\forall~ i \in [1,..,N] \label{eq:joint_control2}
\end{align}
\end{subequations}

\noindent where $\delta^i$ denotes the interference tolerance of agent $i$, i.e., the radio-jamming interference received by agent $i$ from all other agents $j$ should be less than $\delta^i$ in order for agent $i$ to remain operational. 

\subsection{Distributed Tracking-and-Jamming Control}
The joint optimization problem of Eqs. (\ref{eq:joint_control1})-(\ref{eq:joint_control2}) is a hard combinatorial problem which quickly becomes computationally intractable as the number of agents and the number of control actions increases. 

For this reason, in order to tackle this problem and achieve the desired system behavior in real-time, a distributed sub-optimal approach is proposed based on the greedy randomized adaptive search procedure or GRASP \cite{Feo1995,Resende2019}. In essence, GRASP is an iterative randomized sampling technique which operates in two steps. In the first step the algorithm constructs an \textit{initial greedy randomized solution} which is further refined through \textit{local search} in the second step. The algorithm alternates between these two steps, while keeping track of the best solution, until the stopping criteria are met (e.g., usually the number of iterations). The two steps of GRASP are implemented as follows:

\subsubsection{Greedy Randomized Solution}
An initial randomized solution is achieved via random sampling to find the joint control actions $(\hat{u}^j_t,\hat{l}^j_t), \forall j$ with the following steps:
\begin{itemize}
    \item  (a) agent $j,~ \forall j \in [1,..,N]$ receives  $ u^{i \ne j}_{t-1}, \tilde{x}^{i \ne j}_{t} ~\forall i \in [1,..,N]$ at time $t$ and using Eq. (\ref{eq:controlVectors}) computes the set of admissible mobility control actions for all agents $i \ne j \in [1,..,N]$ to create $N$ sets  $ \mathbb{U}^{1}_{t},...,\mathbb{U}^{N}_{t}$, including its own $ \mathbb{U}^{j}_{t}$.
    \item (b) agent $j,~ \forall j \in [1,..,N]$ then locally computes $N$ sets of hypothesized joint mobility and transmit power-level control actions one for each agent $i \ne j \in [1,..,N]$ plus its own i.e., $\left[ \{\mathbb{U}^{1}_{t} \times \mathbb{L}^{1}_{t}\}^1,...,\{\mathbb{U}^{N}_{t} \times \mathbb{L}^{N}_{t}\}^N \right]^j$ denoted as $\left[ \mathbb{Q}^{1}_{t},...,\mathbb{Q}^{N}_{t}\right]^j$ hereafter. The assumption here is that the agents have the same capabilities, i.e., the agents exhibit the same number of mobility control actions and transmit power-levels. This assumption however, is not strict and can be relaxed depending on the application scenario. 
    \item (c) agent $j,~ \forall j \in [1,..,N]$ samples uniformly at random with replacement $n_s$ samples from their local set $\left[ \mathbb{Q}^{1}_{t},...,\mathbb{Q}^{N}_{t}\right]^j$ which are used to create $n_s$ joint control vectors $\omega^j_i = \left[(u^1_t,l^1_t)^i,...(u^j_t,l^j_t)^i,...,(u^N_t,l^N_t)^i   \right]^j, i \in [1,..n_s]$.
    \item (d) Finally, each agent $j$ then solves the optimization problem in Eqs. (\ref{eq:joint_control1})-(\ref{eq:joint_control2}) by searching through $\omega^j_i$ to find an initial global solution to the problem denoted as $\left[(\tilde{u}^1_t,\tilde{l}^1_t),...,(\tilde{u}^j_t,\tilde{l}^j_t)...,(\tilde{u}^N_t,\tilde{l}^N_t)  \right]^j$.
\end{itemize}

\subsubsection{Local Search}
Once an initial global solution is found, a local search is performed trying to find an improvement. To do so, each agent $j,~ \forall j \in [1,..,N]$ performs locally an exhaustive search around the neighborhood of $\left[(\tilde{u}^1_t,\tilde{l}^1_t),...,(\tilde{u}^j_t,\tilde{l}^j_t)...,(\tilde{u}^N_t,\tilde{l}^N_t)  \right]^j$ which is defined as follows: Let $\kappa^{j}$ = $\left[ (\tilde{u}^{j-}_t,\tilde{u}^j_t,\tilde{u}^{j+}_t) \times (\tilde{l}^{j-}_t,\tilde{l}^j_t,\tilde{l}^{j+}_t) \right]$ be the joint combinations of mobility and power-level controls of agent $j$, where notation $x^{+},x^{-}$ denotes two different control actions which are close to $x$, and which are chosen at random from a list of close proximity control actions.

 The neighborhood of $\left[(\tilde{u}^1_t,\tilde{l}^1_t),...,(\tilde{u}^j_t,\tilde{l}^j_t)...,(\tilde{u}^N_t,\tilde{l}^N_t)  \right]^j$ is then defined as the joint combinations of the control vector $[(\tilde{u}^i_t,\tilde{l}^i_t), \forall i \ne j]$ with $\kappa^j$. In essence, agent $j$ performs a local search by changing only its own controls and keeping constant the controls assigned to the other agents. Thus, agent $j$ searches the joint control combinations given by $\left[(\tilde{u}^1_t,\tilde{l}^1_t)\times....\times\kappa^{j}\times...\times(\tilde{u}^N_t,\tilde{l}^N_t) \right]^j$ (where the single pair $(\tilde{u}^j_t,\tilde{l}^j_t)$ is replaced with the vector of pairs $\kappa^j$). Each agent $j$ exhaustively searches the joint combinations in its neighborhood to find the best solution by solving the problem in Eqs. (\ref{eq:joint_control1})-(\ref{eq:joint_control2}).

\begin{figure*}
	\centering
	\includegraphics[width=\textwidth]{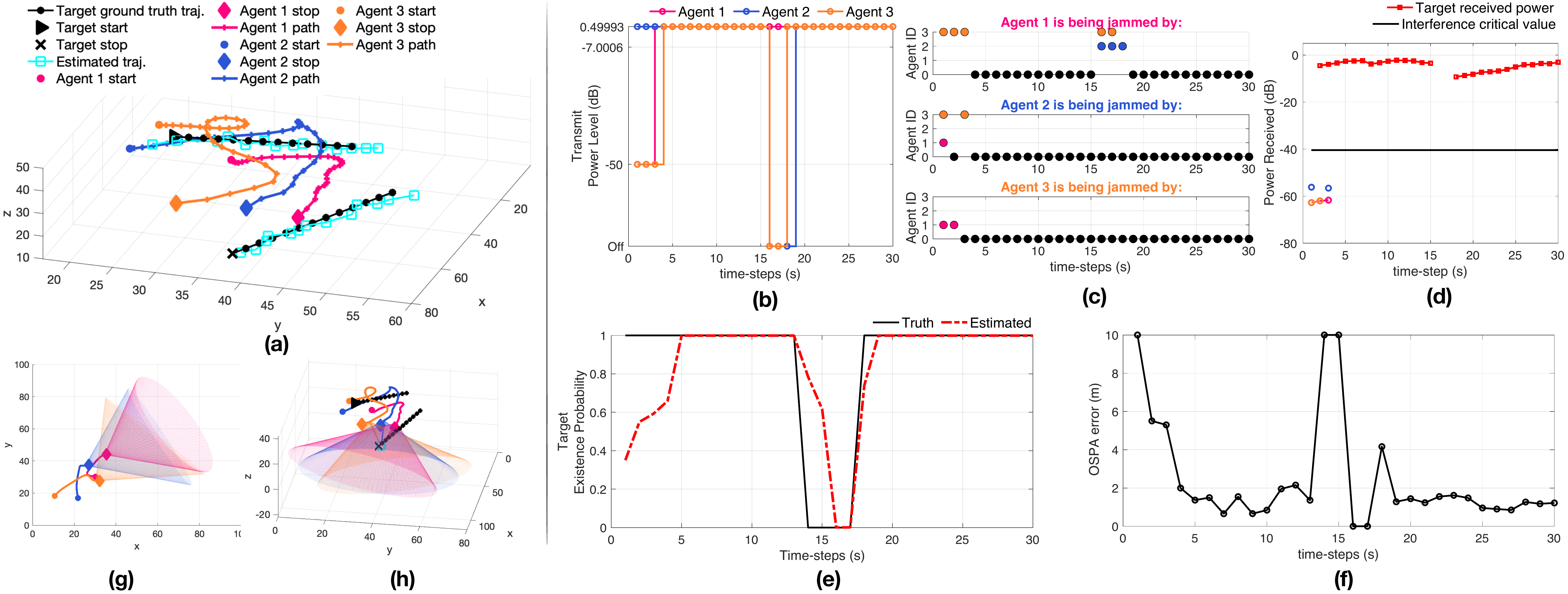}
	\caption{A simulated tracking-and-jamming scenario, where a team of 3 pursuer agents track-and-jam a target.}	
	\label{fig:res1}
	\vspace{-5mm}
\end{figure*}

\subsubsection{Fusion}
Each agent alternates between steps 1) and 2) and the best known feasible solution over all $n_I$ iterations is kept as the final solution. The agents then exchange their local best solutions to converge to the best global solution $(\hat{u}^j_t,\hat{l}^j_t), \forall j$. The best global mobility and transmit power-level control actions are then applied to the agents, which move to their new states, receive target measurements, and apply the update step of the filtering algorithm discussed in Section \ref{ssec:state_estimation} to compute the posterior existence probability $\epsilon_{t}$ and spatial density $s_t$ of the target. 

In this work, a distributed optimization approach based on GRASP is proposed, for finding the joint control actions for the team of agents. Nevertheless, the reader should note that a solution to this problem can also be obtained with other distributed combinatorial optimization techniques including methods based on swarm particle optimization, genetic algorithms, and others. A comparison between different optimization techniques for this problem is left as future work.

\section{Evaluation}\label{sec:Evaluation}

\subsection{Simulation Setup} \label{ssec:sim_setup}
The simulation setup used to evaluate the performance of the proposed approach is as follows: The agents and the target maneuver inside a bounded 3D surveillance area of size $100\text{m} \times 100\text{m} \times 100\text{m}$. The dynamics of the target are according to Eq. \eqref{eq:target_dynamics} with $\Sigma_\text{v}=0.5  \text{I}_3 ~\text{m}/\text{s}^2$ and $\Delta T = 1$s. The dynamics of the pursuer agent are according to Eq. \eqref{eq:controlVectors} with $\Delta_R = [1, 3, 5]$m, $N_\phi = 8$, and $N_\theta= 8$, unless otherwise specified. The pursuer agent detection and jamming model is according to Eq. (\ref{eq:sensing_model}) with $p_D^{\text{max}} = 0.95$, and transmit power levels $\mathbb{L} = [\text{off},-50,-7,0.5]$dB, i.e., $l^j_\text{max} = 0.5$dB $~\forall j$. The path-loss exponent $n_e=2$ and $R_0 = 6$m. The agent measurement model has $\Sigma_\text{w}= \text{diag}([0.8\text{m}, \pi/50 \text{rad}, \pi/50 \text{rad}])$ and clutter rate $\lambda_c = 3$. The conic sensing profile $S$ of each agent has $h_a=40$m and $\theta_a=80$deg and the interference value for each agent $\delta^j$ is set at $-40$dB. The Bernoulli filter parameters are $p_b = 0.02$, $p_s = 0.98$ and the birth density $b_t(x)$ is uniform inside the surveillance space. Finally, the tracking-and-jamming objective function used in the experiments is given by Eq. (\ref{eq:joint_control1}) as $\Xi_2^N$ with $N=3$ and the jamming constraints are given by Eq. (\ref{eq:joint_control2}). To handle the non-linear measurement model, the Bernoulli filter was implemented as a particle filter \cite{Ristic2013}. Finally, in the implementation of the GRASP-based optimization $n_s = 10^4$ samples and $n_I = 100$ iterations were used.

\subsection{Results}
The first experiment, shown in Fig. \ref{fig:res1}, aims to demonstrate the overall behavior of the proposed approach. More specifically, in this simulated scenario which takes place over 30 time-steps, a team of $N=3$ pursuer agents cooperate in order to track-and-jam the target. In particular, Fig. \ref{fig:res1}(a) shows the trajectories of the agents (i.e., blue, purple and orange) while tracking and jamming the target (i.e., ground truth is shown with black line, estimated is shown with cyan line). The agents are initially located at the $(x,y,z)$ coordinates $[30,30,30]^\top, [20,10,10]^\top, [10,15,15]^\top$ and the target is spawned inside their sensing range with initial state $[20, 20, 20, 1.5, 2, 1.7]^\top$. As shown in the figure, the agents maneuver around the target in such a way so that (a) the tracking-and-jamming performance is maximized while (b) the jamming interference amongst the team is kept below the critical value of -40dB. More specifically, 
Fig. \ref{fig:res1}(b) shows the transmit power level of each agent during this experiment and  Fig. \ref{fig:res1}(c) shows the jamming agents (i.e., due to jamming interference). For instance, agent 1 is being jammed by agent 3 during time-steps 1 to 3 and then again during time-steps 16 and 17. Additionally, during time-steps 16 to 18 agent 1 is also being jammed by agent 2 as shown in the first row of Fig. \ref{fig:res1}(c). Then, Fig. \ref{fig:res1}(d) shows the target received power (i.e., red line) and the agent jamming interference in dB indicated by the colored circles. Figure \ref{fig:res1}(e) shows the estimated target existence probability and Fig. \ref{fig:res1}(f) shows the tracking error i.e., the optimal sub-pattern assignment (OSPA) metric \cite{Schuhmacher2008c} of order 2 with cut-off value of 10m i.e., OSPA(p=2,c=10). In this experiment the target is occluded between time-steps 14 to 17 and thus from the agents' perspective the target seems to disappear at time-step 14 and then to re-appear at time-step 18 as shown in Fig. \ref{fig:res1}(e). The agents try to estimate the target's appearance and disappearance events captured by the existence probability (i.e., red line) in Fig. \ref{fig:res1}(e). 

From the results it is also observed that during the first few time-steps the agents cause interference to each other. For this reason, the agents transmit at lower power levels to accommodate this interference as shown in Figs. \ref{fig:res1}(b)-(c). Moreover, at time-step 16, agent 1 is being jammed by agent 2 and agent 3 as shown in Fig. \ref{fig:res1}(c). Subsequently, agents 2 and 3 switch off their antennas as shown in Fig. \ref{fig:res1}(b) in order to respect the interference limit of agent 1. Figure \ref{fig:res1}(d) verifies that at time-step 16 agent 1 does not receive any power from agents 2 and 3. Figure \ref{fig:res1}(g) shows the configuration of the agents' sensing profile at this particular time step and as it can be seen, agent 1 resides inside the jamming range of agents 2 and 3. Then, the spike in the tracking error at time-step 15 is due to a cardinality estimation error, i.e., the agents believe that the target is present at time-step 15, however in reality the target at time-step 15 is absent. This error, however, is corrected in the next time-step. Finally, Fig. \ref{fig:res1}(h) shows the configuration of the agents at time-step 30 along with their antenna orientations. As illustrated, the agents take positions which avoid interferences with each other, while at the same time the positions taken by the agents maximize the target received power. 

\begin{figure}
	\centering
	\includegraphics[width=\columnwidth]{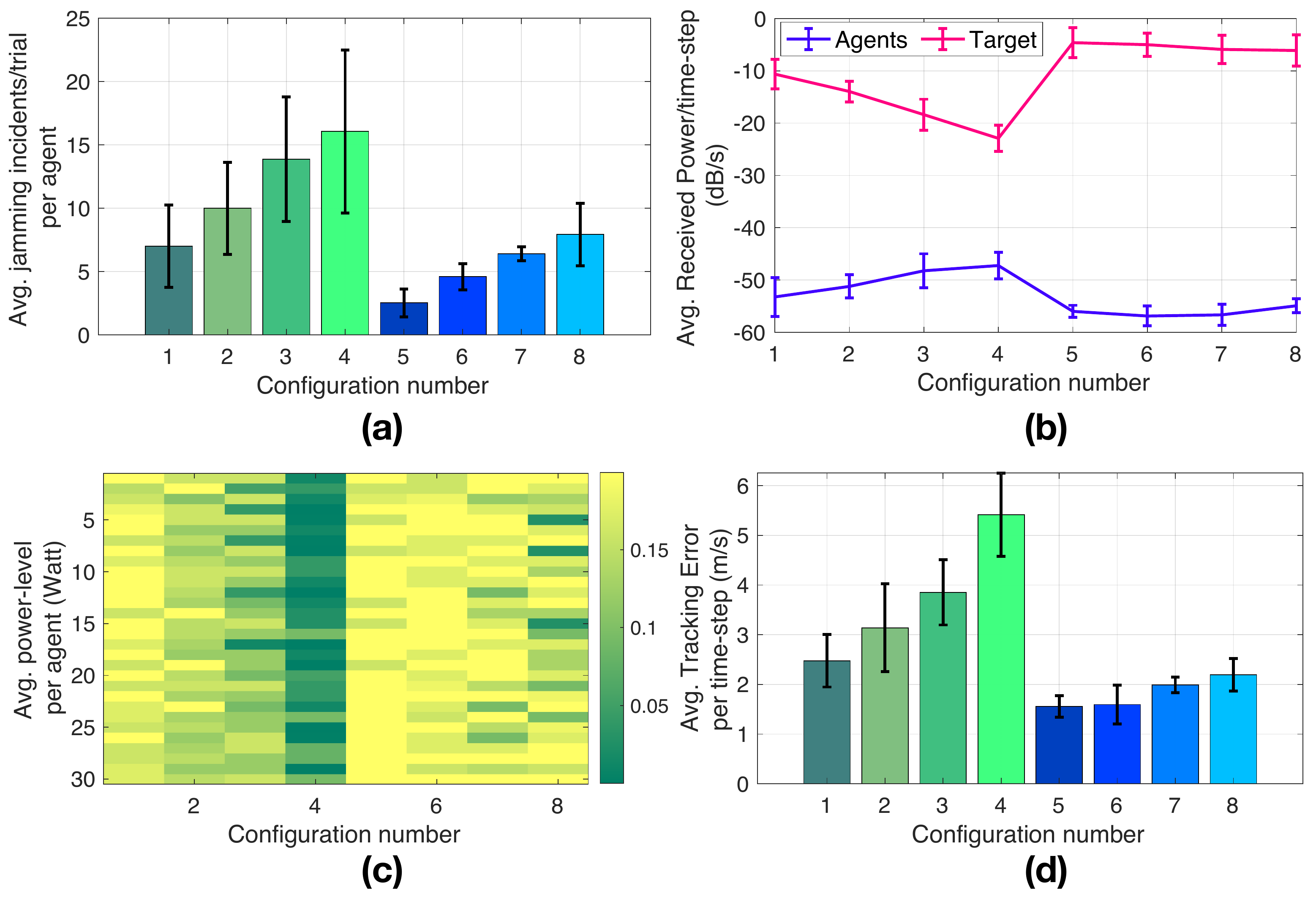}
	\caption{Impact of the number of mobility controls and the size of the opening angle $\theta_a$ of sensing profile $\mathcal{S}$ on the performance of the proposed approach.}	
	\label{fig:res2}
	\vspace{-3mm}
\end{figure}

The second experiment aims to investigate in more detail the impact of the various parameters on the behavior of the proposed approach. In particular, it investigates the impact of the opening angle $\theta_a$ of the sensing profile $S$ and the number of total mobility control actions on the performance of the system. 
Intuitively, it is expected that as $\theta_a$ increases, the sensing range increases, and thus the number of times the agents interfere with each other increases as well. However, as the system's degrees of freedom, i.e., the number of admissible mobility control actions, increase, additional options are provided for the agents to try to avoid the interference.

\begin{table} \label{table:1}
\caption{Parameter Configurations}
\renewcommand{\arraystretch}{0.8}
\begin{center}
	\begin{tabular}{| c | c | c |}
		\hline
		Config. \# & \# Mob. Controls & $\theta_a$ (deg)\\
		\hline \hline
		1 & 11 & 60\\  \hline
		2 & 11 & 80\\  \hline
		3 & 11 & 100\\ \hline
		4 & 11 & 120\\ \hline
		5 & 79 & 60\\  \hline
	    6 & 79 & 80\\  \hline
        7 & 79 & 100\\  \hline
        8 & 79 & 120\\
	    \hline
	\end{tabular}
\end{center}
\vspace{-7mm}
\end{table}

To verify this assumption the parameter configurations shown in Table I were used. This table shows that for configurations 1-4 each agent has a total of 11 mobility control actions, whereas configurations 5-8 allow for a total of 79  mobility actions per agent. Moreover, the opening angle $\theta_a$ takes increasing values from 60deg to 120deg. The experimental setup is as follows (with all the other parameters taking values according to  Section \ref{ssec:sim_setup}): First the target is randomly spawned within the surveillance area. Then, the initial locations of 3 purser agents are randomly sampled, from within a sphere with radius 20m centered at the target location. The antennas of the agents point to the center of the sphere. For each configuration shown in Table I the system runs for 30 time-steps (i.e., one trial). This process is performed for each configuration 20 times and the results are averaged and shown in Fig. \ref{fig:res2}. More specifically, Fig. \ref{fig:res2}(a) shows the average jamming incidents per trial per agent for each configuration, Fig. \ref{fig:res2}(b) shows the average received power for the target and agents, Fig. \ref{fig:res2}(c) shows the average transmit power-level in Watts at each time-step for the 8 configurations, and finally Fig. \ref{fig:res2}(d) shows the average tracking error. It can be observed from these results that with 11 mobility control actions as the opening angle $\theta_a$ becomes larger the number of times that the agents are jamming each other increases and so does the agent received power due to interference as shown in Figs. \ref{fig:res2}(a)-(b). This is due to the limited number of admissible control actions which prohibit the agents to acquire positions which are interference free. In addition, since the target detection is linked to the transmit power-level (as shown in Eq. (\ref{eq:sensing_model})), the agents by turning off their antennas put themselves in a disadvantage regarding their tracking performance which can result in complete tracking failure. To overcome this issue, the agents try to transmit at low power-levels (i.e., Fig. \ref{fig:res2}(c)). This however, not only causes reduced tracking performance (i.e., Fig. \ref{fig:res2}(d)) but also reduced target jamming performance (i.e., Fig. \ref{fig:res2}(b)). On the other hand, once the system's degrees of freedom are increased, large opening angles can be handled more efficiently with improved jamming and tracking performance and reduced interference as illustrated in Fig. \ref{fig:res2} for configurations 5-8.

\begin{figure}
	\centering
	\includegraphics[width=\columnwidth]{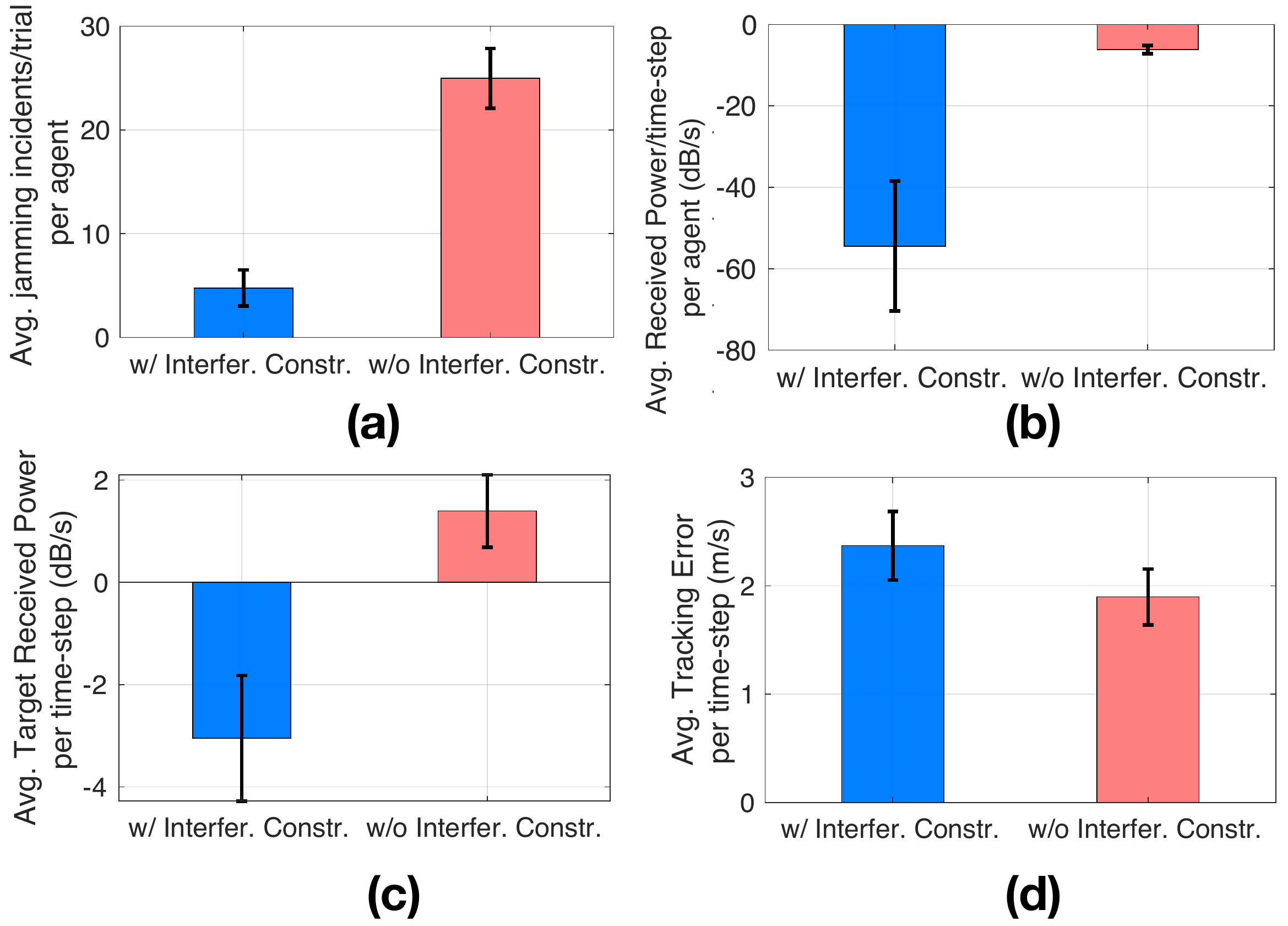}
	\caption{Impact of the interference constraints i.e., Eq. (\ref{eq:joint_control2}), on the tracking and jamming performance.}
	\label{fig:res3}
	\vspace{-3mm}
\end{figure}

The last experiment aims to investigate how the interference constraints in Eq. (\ref{eq:joint_control2}) affect the tracking and jamming performance of the system. Again, the same simulation setup discussed in the previous paragraph is followed, where a target is randomly spawned inside the surveillance area and then within a sphere around the target location 3 agents are spawned. The system runs for trials with duration 30 time-steps and evaluated with the interference constraints enabled and disabled. Twenty (20) trials are conducted for each case with the simulation parameters and values set according to Section \ref{ssec:sim_setup}. The average results are shown in Fig. \ref{fig:res3}. As expected, by disabling the interference constraints the number of jamming incidents per agent increases as shown in Fig. \ref{fig:res3}(a). As a result, the average received power per agent also increases as shown by Fig. \ref{fig:res3}(b) significantly above the critical value, and thus the system is driven into failure. On the other hand, it is shown in Fig. \ref{fig:res3}(c) that the target received power is higher with the interference constraints turned off. The results show that when the interference constraints are disabled the agents can get much closer to the target and to each other. As a result, the target receives much higher power compared to the power received with the interference constraints enabled. Similarly, with the interference constraints turned off, the agents optimize the joint target detection probability of Eq. \eqref{eq:joint_control1} without any constraints and this results in better tracking performance as shown in Fig. \ref{fig:res3}(d). However, from the aforementioned results it is clear that the interference constraints are vital for keeping the system operational at all times while achieving a satisfactory tracking and jamming performance as shown in Fig. \ref{fig:res3}.

\section{Conclusion} \label{sec:Conclusion}
This work investigated the problem of cooperative target tracking and target radio-jamming with a team of pursuer agents. A novel distributed control framework is presented, in which a team of pursuer agents select their mobility and transmit power-level control actions that result in accurate target tracking and  uninterrupted target radio-jamming, while avoiding the jamming interference amongst them. Future directions include a real-world implementation of the proposed system and its extension to multiple targets.

\bibliographystyle{IEEEtran}
\bibliography{IEEEabrv,main}

\end{document}